\documentclass[doublecol]{epl2} 
\usepackage[table]{xcolor}

\begin{document}

\title{Google matrix analysis of C.elegans neural network}

\author{
Vivek\ Kandiah\inst{1}, 
Dima L.\ Shepelyansky\inst{1}} 
\shortauthor{V.Kandiah and D.L.Shepelyansky} 
\shorttitle{Google matrix of C.elegans neural network}
\institute{
 \inst{1} Laboratoire de Physique Th\'eorique du CNRS (IRSAMC), 
  Universit\'e de Toulouse, UPS, F-31062 Toulouse, France
}
%\date{\today} 

\abstract{
We study the structural properties of the neural network of 
the C.elegans (worm) from a directed graph point of view. 
The Google matrix analysis is used to characterize 
the neuron connectivity structure and node classifications 
are discussed and compared with physiological properties of the cells. 
Our results are obtained by a proper definition of neural directed network and 
subsequent eigenvector analysis which recovers some results of 
previous studies. Our analysis highlights 
particular sets of important neurons 
constituting the core of the neural system. 
The applications of PageRank, CheiRank and 
ImpactRank to characterization of interdependency of neurons are
discussed.
}
\pacs{84.35.+i}{Neural networks}
\pacs{89.75.Hc}{Networks and genealogical trees}
\pacs{89.20.Hh}{World Wide Web, Internet}
%\pacs{89.75.Da}{Systems obeying scaling laws}
%\pacs{89.75.Fb}{Structures and organization in complex systems}
%\pacs{89.20.-a}{Interdisciplinary applications of physics}

\date{November 8, 2013}

\maketitle

%%%%%%%%%%%%%%%%%%%%%%%%%%%%%%%%%%%%%%%%%%%%%%%%%%%%%%%%%%%%%%%%%%%%%%%%%%%%%%%
%******************************************************************************
\section{Introduction}
%******************************************************************************
%%%%%%%%%%%%%%%%%%%%%%%%%%%%%%%%%%%%%%%%%%%%%%%%%%%%%%%%%%%%%%%%%%%%%%%%%%%%%%%
The human brain neural network has an enormous complexity
containing about $10^{11}$ neurons and $10^{14}$  synapses linking various neurons
\cite{neuronumber}. Such a complex network can only be  compared 
with the World Wide Web (WWWW) which indexed size is
estimated to be of about $10^{10}$ pages \cite{wwwsize}.
This comparison gives an idea that the methods of computer science,
developed for WWW analysis, can be suitable for
the investigations of neural networks.
Among these methods the PageRank algorithm of the Google matrix
of WWW \cite{brin} clearly demonstrated its efficiency
being at the heart of Google search engine \cite{googlebook}.
Thus we can expect that the Google matrix analysis can find useful
applications for the neural networks.
This approach has been tested in \cite{szbrain}
on a reduced brain model  of mammalian thalamocortical systems
studied in \cite{izhikevich}. However, it is more interesting to 
perform the Google matrix analysis for real neural networks. 
In this Letter we apply this analysis to characterize the properties of 
neural network of {\it  C.elegans} (worm). 
The full connectivity of this directed network
is known and documented at \cite{wormatlas}. The number of 
linked neurons (nodes)
is $N=279$ with the number of 
synaptic connections and gap junctions 
(links) between them being $N_{\ell}=2990$.
%Emma has $ 2287$. 

Recently, there is a growing interest to the 
complex network approach for investigation of brain neural networks 
\cite{arenas,bullmore},\cite{kaiser,chklovskii},\cite{towlson}.
Generally these networks are directional but 
it is difficult to determine directionality of links
by physical and physiological measurements. 
Thus, at present, the worm network is 
practically the only example of neural network where the
directionality of all links is established \cite{wormatlas}.
The analysis of certain properties this directed network  
has been reported recently in \cite{chklovskii,towlson},
however, the approach based on the Google matrix has
not been used yet. Thus we think that this 
study will allow to highlight the features of 
worm network using recent advancements of computer science.

\section{Google matrix construction}
The Google matrix $G$ of {\it C.elegans} is constructed using 
the connectivity matrix elements $S_{ij}=S_{syn,ij}+S_{gap,ij}$,
where $S_{syn}$ is an asymmetric matrix of synaptic links whose elements are $1$ 
if neuron $j$ connects to neuron $i$ through 
a chemical synaptic connection and $0$ otherwise. 
The matrix part $S_{gap}$ is a symmetric matrix describing gap junctions 
between pairs of cells, $S_{gap,ij}=S_{gap,ji}=1$ if neurons $i$ and $j$ 
are connected through a gap junction and $0$ otherwise.
Following the standard rule \cite{brin,googlebook},
the matrix elements $S_{ij}$ are renormalized
($S_{ij} = S_{ij}/\sum_i S_{ij}$)
for each column with non-zero elements; the columns with all zero elements   
are replaced by columns with all elements  $1/N$. 
Thus the sum of elements in each column is equal to unity
and the Google matrix takes the form 
\begin{equation}
G_{ij}=\alpha S_{ij} + (1-\alpha)/N \; \; .
\label{eq1}
\end{equation}
Here $\alpha$ is the damping factor introduced in \cite{brin}. 
In the context of the WWW, the last term of the equation describes 
a probability for a random surfer to jump on 
any node of the network \cite{googlebook}.
We use the usual value $\alpha=0.85$ \cite{googlebook}.
All matrix elements $S_{syn,ij}, S_{gap,ij}, S_{ij}$ are given at
\cite{wormgooglematrix}

\begin{figure}[!ht] 
\begin{center} 
\includegraphics[width=7.7cm]{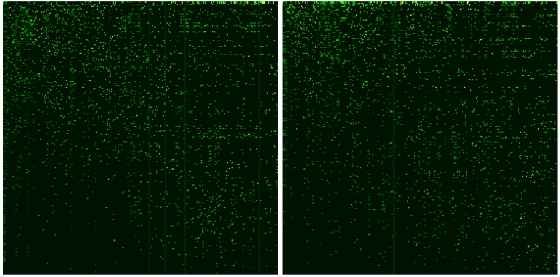}
\caption {(Colour on-line)
Google matrix $G$ (left) and $G^*$ (right) for the neural network 
of {\it C.elegans} for $N=279$ connected neurons. 
Matrix elements $G_{KK'}$ are shown in the basis 
of PageRank index $K$ (and $K'$)
and elements $G^*_{K^*,K^{*'}}$ are shown in the basis of
CheiRank index $K^*$ (and $K^{*'}$)
at $\alpha=0.85$. Here, $x$ and $y$ axes show $1 \leq K, K' \leq N$ and 
$1 \leq K^*, K^{*'} \leq N$; the elements $G_{11}, {G^*}_{11}$  
are placed at the top left corner; color is proportional to
the square root of matrix elements which are changing 
from black at minimum value $(1-\alpha)/N$ to light yellow at maximum.}
\label{fig1}
\end{center}
\end{figure}

The eigenspectrum $\lambda_i$ and right eigenvectors $\psi_i(j)$ of $G$
satisfy the equation  
\begin{equation}
 \sum_{j'}  G_{jj'} \psi_i(j') = \lambda_i \psi_i(j) \; .
\label{eq2} 
\end{equation} 
The eigenvector at $\lambda=1$ is known as the PageRank vector.
According to the Perron-Frobenius theorem \cite{googlebook}
its elements $P(j) \sim \psi_1(j)$ are positive  and 
their sum is normalized to unity. Thus $P(j)$ gives a probability to find
a random surfer on a node $j$. 
All nodes can be ordered in a decreasing 
order of probability $P(K_j)$ with highest probability 
at top values of PageRank index $K_j=1, 2, ...$.
For large matrices $P(j)$ can be found numerically by the iteration method
\cite{googlebook} but for {\it C.elegans} case it can be obtained by
a direct matrix diagonalization.

It is also useful to consider the Google matrix obtained from the network
with inverted directions of links (see e.g. \cite{alik,zzs},\cite{2dmotor}). 
The matrix $G^*$
for this network with inverted direction of links is constructed 
following the same definition (\ref{eq1}). 
The PageRank vector of this matrix $G^*$ is called the 
CheiRank vector with probability  $P^*(K^*_j)$ 
and CheiRank index $K^*$. According to the known results \cite{brin,googlebook}
the top nodes of PageRank are the most popular pages, while the top
nodes of CheiRank are the most communicative nodes \cite{zzs,2dmotor}.

\begin{figure}[!ht] 
\begin{center} 
\includegraphics[trim = 0mm 0mm 0mm 0mm, clip, width=7.7cm]{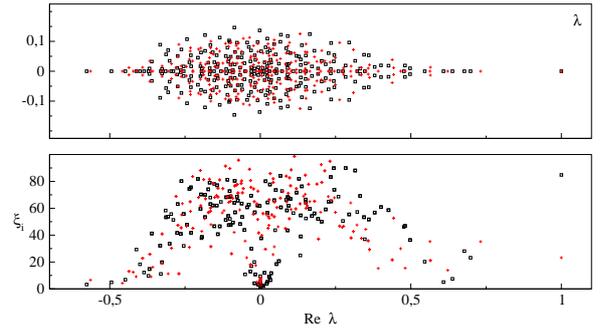}
\caption {(Colour on-line)
Top panel: spectrum of eigenvalues $\lambda$ for the Google matrices 
$G$ and $G^*$ at $\alpha=0.85$ (black and red symbols). Bottom panel: 
IPR $\xi$ of eigenvectors as a function of corresponding $Re \lambda$
(same colors).}
\label{fig2}
\end{center}
\end{figure}

The structure of the matrix elements 
of $G$, presented in the PageRank ordering of nodes,
and $G^*$, presented in the CheiRank ordering of nodes,
is shown in Fig.~\ref{fig1}. The number of nonzero elements $N_G$
with indexes less than $K$ is respectively
$N_G/K \approx 1.2, 10$ at $K=10, 100$.
These values correspond approximately 
to those of  WWW networks of
Universities of Cambridge and Oxford  being
significantly smaller than the values of Twitter network 
characterized by a strong connectivity between top PageRank nodes
with $N_g/K \approx 100$ for $K=100$
(see Fig.2 in \cite{physrev}). 
We note that the average number of links per neuron
is $\eta = N_{\ell}/N =10.71$ being approximately the same
as for WWW of Universities of Cambridge 
and Oxford in 2006 \cite{2dmotor}.

The global matrix structure
is asymmetric.  This leads to a complex
spectrum of eigenvalues of $G$ and $G^*$ as shown in 
top panel of Fig.~\ref{fig2}. 
The imaginary part of eigenvalues 
is distributed in a range
$-0.2 < Im \lambda <0.2$ which is 
more narrow than for the networks
of Wikipedia and UK universities \cite{wikispectrum}.
This is related to a significant number of symmetric links.
On the other side the networks of Le Monde
or Python have comparable width for 
$Im \lambda$ \cite{wikispectrum}.
We find that the second by modulus
eigenvalues are $\lambda_2= 0.8214$ for $G$
and $\lambda_2= 0.8608$ for $G^*$.
Thus the network relaxation time 
$\tau = 1/|\ln \lambda_2|$ is approximately
$5, 6.7$ iterations of $G, G^*$. 

The properties of 
eigenstates $\psi_i$ can be characterized by the Inverse Participation Ratio
(IPR) $\xi_i=\left( \sum_j |\psi_i(j)|^2 \right)^2/\sum_j |\psi_i(j)|^4$,
which is broadly used in analysis of electron conductivity
in disordered systems (see e.g. \cite{wikispectrum,physrev}).
This quantity effectively determines the number of
nodes on which is located an eigenstate $\psi_i$.
We see that some eigenstates have rather large
$\xi \approx N/3$ while others have $\xi$ located only on about ten nodes.
We will return to the discussion of properties of eigenstates later.

\section{CheiRank versus PageRank}

\begin{figure}[!ht] 
\begin{center} 
\includegraphics[trim = 0mm 0mm 0mm 0mm, clip,width=7.5cm]{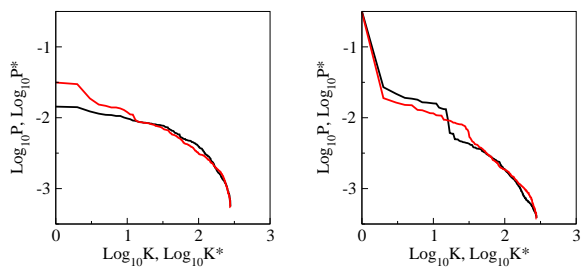}
\caption {(Colour on-line)
\emph{Left panel:} dependence of PageRank (CheiRank) probability 
$P(K)$ ($P^*(K^*)$) on its index $K$ ($K^*$) 
shown by black (red) curve.
\emph{Right panel:} dependence of 
ImpactRank probability $P$ ($P^*$) on its index $K$ ($K^*$),
obtained via propagator of $G$ ($G^*$)
at $\alpha=0.85$ and $\gamma= 0.7$
for the initial probability located on
neuron RMGL (see text).} 
\label{fig3}
\end{center}
\end{figure}

The dependence of probabilities of PageRank and CheiRank vectors
on their indexes $K$ and $K^*$ is shown in Fig.~\ref{fig3}.
A formal fit for a power law dependence
$P \propto 1/K^{\nu}, P^* \propto 1/K^{* \nu}$
in the  range $1 \leq K, K^* \leq 200$
gives $\nu = 0.33  \pm 0.03$ 
for PageRank and $\nu = 0.50 \pm 0.03$ for CheiRank.
Of course, the number of nodes is small compared
to the WWW or Wikipedia networks but on average
we can say that a power law provides 
a satisfactory description of data.
We note that the values of $\nu$ are notably
smaller than the usual exponent value
$\nu \approx 0.9$ (in $K$), $0.6$ (in $K^*$)
found for the WWW or Wikipedia networks
(see e.g. \cite{googlebook,zzs}).
Also in our neural network we find
that the exponent in $K$ is smaller
then in  $K^*$ while usually 
one finds the opposite situation.
Also we have IPR $\xi \approx  85$ for $P$
and $\xi \approx  23$ for $P^*$
so that PageRank is distributed over 
a larger number of neurons.
It is possible that such an inversion is 
related to a significant importance of
outgoing links in neural systems:
in a sense such links transfer orders,
while ingoing links bring instructions 
to a given neuron from other neurons.
We note that somewhat similar 
situation appears for networks
of Business Process Management (BMP)
where {\it Principals} of a company
are located at the top CheiRank position
while the top PageRank positions
belong to company {\it Contacts} \cite{abel}.

The correlations between PageRank and CheiRank vectors
is convenient to characterize
by the correlator $\kappa=N\sum_i P(i) P^*(i) -1=0.125$.
For C.elegans network the value of correlator
is relatively small compared to those 
found for Wikipedia ($\kappa \approx 4$)
and WWW of UK universities ($\kappa \sim 3$) \cite{2dmotor}.
In a sense for C.elegans neural network
the situation if more similar to the networks 
of Linux Kernel  ($\kappa \approx 0$) 
\cite{alik} and BMP ($\kappa =0.164$) \cite{abel}.
Thus, the C.elegans network has practically no correlations
between ingoing and outgoing links.
It is argued in \cite{alik,2dmotor} that such a  network structure
allows to perform a control of information
flow in a more efficient way.
Namely, it allows to reduce the propagation of 
errors in software codes. It seems that the neural networks
also adopt such a structure.

\begin{figure}[!ht] 
\begin{center} 
\includegraphics[width=7.5cm]{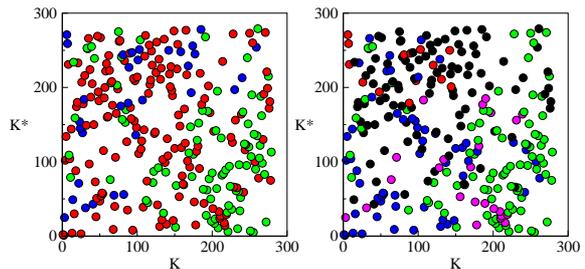}
\caption {(Colour on-line)
PageRank - CheiRank plane  
$(K,K^*)$ showing distribution of neurons 
according to their ranking. 
\emph{Left panel :} soma region coloration - head (red), middle (green), 
tail (blue). \emph{Right panel : } neuron type coloration - 
sensory (red), motor (green), interneuron (blue), 
polymodal (purple) and unknown (black). 
The classifications and colors 
are given according to WormAtlas \cite{wormatlas}.}
\label{fig4}
\end{center}
\end{figure}

Each neuron $i$  belongs to two ranks $K_i$ and $K^*_i$
and it is convenient to represent the distribution of neurons
on the two-dimensional plane (2D)
of PageRank-CheiRank indexes $(K,K^*)$  shown in Fig.~\ref{fig4}.
The plot confirms that there are little correlations between both ranks
since the points are scattered over the whole plane. 
Neurons ranked at top $K$ positions of PageRank
have their soma located mainly 
in both extremities of the worm (head and tail) 
showing that neurons in those regions 
have important connections coming from many other neurons
which control head and tail movements. 
This tendency is even more visible 
for neurons at top $K^*$ positions of CheiRank 
but with a preference for head and middle regions. 
In general, neurons, that have their soma 
in the middle region of the worm,
are quite highly ranked in CheiRank but not in PageRank. 
The neurons located at the head region
have top positions in CheiRank and also 
PageRank, while the middle region
has some top CheiRank indexes but 
rather large indexes of PagRank
(Fig.~\ref{fig4} left panel).
The neuron type coloration 
(Fig.~\ref{fig4} right panel) also reveals 
that sensory neurons are at top PageRank positions
but at rather large CheiRank indexes, 
whereas in general motor neurons are in the opposite situation.

\begin{table}[!ht]
\caption{Top twenty neurons of PageRank (PR), CheiRank (CR); 
ImpactRank of $G$ (IMPR) 
and $G^*$ (IMCR) at initial state RMGL at $\gamma=0.7$; 
following \cite{wormatlas}, the colors mark: 
interneurons (blue \emph{bu}), 
motor neurons (green \emph{gn}), sensory neurons (red \emph{rd}), 
polymodal neurons (purple  \emph{pu}).}
\begin{center}
\resizebox{0.9\columnwidth}{!}{
\begin{tabular}{|c|c|c||c|c|} 
\hline 
 & PR & CR & IMPR & IMCR \\ 
\hline 
\hline 
1 & AVAR \emph{(bu)} & AVAL \emph{(bu)} & RMGL \emph{(bu)}   & RMGL \emph{(bu)}\\
2 & AVAL \emph{(bu)} & AVAR \emph{(bu)} & URXL \emph{(bu)}   & AVAL \emph{(bu)}\\
3 & PVCR \emph{(bu)} & AVBR \emph{(bu)} & ADEL \emph{(rd)}    & ASHL \emph{(rd)}\\
4 & RIH  \emph{(bu)} & AVBL \emph{(bu)} & AIAL \emph{(bu)}   & AVBR \emph{(bu)}\\
5 & AIAL \emph{(bu)} & DD02 \emph{(gn)}& IL2L \emph{(rd)}    & URXL \emph{(bu)}\\ 
6 & PHAL \emph{(rd)}  & VD02 \emph{(gn)}& ADLL \emph{(rd)}    & AVEL \emph{(bu)}\\
7 & PHAR \emph{(rd)}  & DD01 \emph{(gn)}& PVQL \emph{(bu)}   & RIBL \emph{(bu)}\\
8 & ADEL \emph{(rd)}  & RIBL \emph{(bu)} & ALML \emph{(rd)}    & RMDR \emph{(pu)}\\
9 & HSNR \emph{(gn)}& RIBR \emph{(bu)} & ASKL \emph{(rd)}    & RMDL \emph{(pu)}\\
10 & RMGR\emph{(bu)} & VD04 \emph{(gn)}& CEPDL\emph{(rd)}    & RMDVL \emph{(pu)}\\
11 & VC03\emph{(gn)}& VD03 \emph{(gn)}& ASHL \emph{(rd)}    & AVAR \emph{(bu)}\\
12 & AIAR\emph{(bu)} & VD01 \emph{(gn)}& AWBL \emph{(rd)}    & SIBVR\emph{(bu)}\\
13 & AVBL\emph{(bu)} & AVER \emph{(bu)} & SAADR\emph{(bu)}   & AIBR \emph{(bu)}\\
14 & PVPL\emph{(bu)} & RMEV \emph{(gn)}& RMHR \emph{(gn)}  & ADAL \emph{(bu)}\\
15 & AVM \emph{(rd)}  & RMDVR\emph{(pu)} & RMHL \emph{(gn)}  & RMHL \emph{(gn)}\\
16 & AVKL\emph{(bu)} & AVEL \emph{(bu)} & RIH  \emph{(bu)}   & AVBL \emph{(bu)}\\
17 & HSNL\emph{(gn)}& VD05 \emph{(gn)}& OLQVL\emph{(pu)}& SIBVL\emph{(bu)}\\
18 & RMGL\emph{(bu)} & SMDDR\emph{(pu)} & AIML \emph{(bu)}   & ASKL \emph{(rd)}\\
19 & AVHR\emph{(bu)} & DD03 \emph{(gn)}& HSNL \emph{(gn)}  & RID \emph{(bu)}\\
20 & AVFL\emph{(bu)} & VA02 \emph{(gn)}& SDQR \emph{(bu)}   & SMBVL\emph{(pu)}\\
\hline 
\end{tabular}}
\end{center}
\label{table1}
\end{table}

The top $20$ neurons of PageRank and CheiRank vectors 
are given in the first two columns of Table~\ref{table1}. 
We note that both rankings favor important signal 
relaying  neurons such as $AVA$ and $AVB$ 
that integrate signals from crucial nodes 
and in turn pilot other crucial nodes.
Neurons $AVAL,AVAR$, $AVBL,AVBR$ and $AVEL,AVER$ are considered 
to belong to the rich club analyzed in\cite{towlson}.
The right panel in Fig.~\ref{fig3}
and second two columns of Table~\ref{table1}
represent ImpactRank which is discussed below.

\begin{figure}[!ht] 
\begin{center} 
\includegraphics[width=7.5cm]{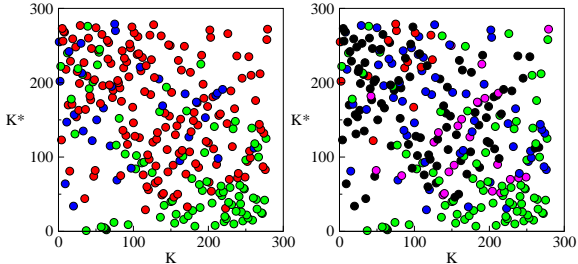}
\caption {(Colour on-line)
Distribution of neurons in the
plane $(K,K^*)$ of equal opportunity ranks (see text);
colors are the same as in Fig.~\ref{fig4}.
}
\label{fig5}
\end{center}
\end{figure}

We can also use 2DRank index $K_2$, discussed in \cite{zzs},
which counts nodes in order of their appearance on 
ribs of squares in $(K,K^*)$ plane with the square size
growing from $K=1$ to $K=N$. The top neurons in $K_2$
are AVAL, AVAR, AVBL, AVBR, PVCR. Thus at the top $K_2$
values we find dominance of interneurons.
More detailed listings are available at \cite{wormgooglematrix}.

It may be also useful to consider
renormalized equal opportunity rank recently discussed in \cite{banky}.
In this approach PageRank probability of node $i$ is replaced by
$P(i)/d(i)$ where $d(i)$ is in-degree of node $i$.
For the Google matrix this recipe should be replaced by
$P(i) \rightarrow P(i)/\sum_j G_{ij}$ and respectively for CheiRank
by  $P^*(i) \rightarrow  P^*(i)/\sum_j G^*_{ij}$.
The corresponding rank indexes $K,K^*$ rank the neurons 
in the decreasing order of these
renormalized probabilities. The distribution of 
nodes in the plane $(K,K^*)$ is shown in Fig.~\ref{fig5}. 
In this ranking the top $K$ nodes
correspond to important sensory neurons
rather than information relaying centers, 
whereas the top nodes of $K^*$
are composed mainly by motor neurons.
Thus such an approach allows to highlight
additional  features  of C.elegans network
being complementary to PageRank and CheiRank
properties discussed above.
Tables for neuron renormalized ranking are available at
 \cite{wormgooglematrix}.

\section{ImpactRank}

In certain cases it is useful to determine an influence or impact
of a given neuron on other neurons. A recent proposal 
of ImpactRank, described in \cite{physrev},
is based on the probability distribution of 
a vector $v_f =(1-\gamma) (1- \gamma G)^{-1} v_0$,
$v^*_f =(1-\gamma) (1- \gamma G^*)^{-1} v_0$,
where $v_0$ is initially populated neuron.
The vector $v_f$ can be viewed as a Green function propagator.
The computation of $v_f$ is obtained numerically by a
summation of geometrical expansion series which
are convergent within approximately first $200$ terms
at $\gamma \sim 0.7$ (see also \cite{physrev}).
The distributions of probabilities of 
ImpactRank $P(i)=v_f(i)$, $P^*(i)=v^*_f(i)$
versus the corresponding ImpactRank indexes $K,K^*$
are shown in Fig.~\ref{fig3} (right panel)
for the initial state neuron $RMGL$.
The corresponding top $20$ ImpactRank neurons 
influenced by $RMGL$ are given in columns
$IMPR$, $IMCR$ of Table~\ref{table1}.
The analysis of neurons linked to $RMGL$ shows that indeed,
ImpactRank correctly selects neurons
influenced by $RMGL$. 
The neurons in the top list
of $P(i)$ are those pointed by 
outgoing links of  $RMGL$
while those in the top list of $P^*(i)$
are those that have ingoing links to
 $RMGL$.
Such a method can be easily applied to
other initial neuron states of interest
showing a contamination propagation
over the neural network starting
from initial neuron $RMGL$.

\section{Properties of Eigenstates}

The Google matrix analysis of the Wikipedia hyperlink network 
\cite{wikispectrum}
demonstrated that the eigenstates with large values of $|\lambda|$
select well defined communities. Thus we can expect that
other eigenstates of matrices $G$ and $G^*$
with $|\lambda| <1$ correspond to certain physiological functions of
worm neural network. 
It is convenient to order index of eigenstates
in a decreasing order of $|\lambda_i|$
with $\lambda_1=1$.

\begin{table}[!ht]
\caption{Top ten neurons of the eigenvectors 
of $G$ (left panel) and  $G^*$ (right panel) 
corresponding to the 10th largest eigenvalues $|\lambda|$; 
IPR are respectively $\xi \approx 5$ and $\xi \approx 4$.}
\begin{center}
\resizebox{7.5cm}{!}{
\begin{tabular}{|c|c|c|} 
\hline 
 & $\lambda_{10}=-0.49446$ & $|\psi_i|$ \\ 
\hline 
\hline 
1 & AIAR & 0.11986 \\ 
2 & AIAL & 0.11159 \\ 
3 & ASIL & 0.096475 \\ 
4 & ASIR & 0.096236 \\ 
5 & AWAR & 0.024228 \\ 
6 & ASHR & 0.022241 \\ 
7 & RMGR & 0.018502 \\ 
8 & AIMR & 0.018387 \\ 
9 & ADLL & 0.01837 \\ 
10 & PVQL & 0.017547 \\ 
%11 & PVQR & 0.01742 \\ 
%12 & ASEL & 0.017256 \\ 
%13 & ASGR & 0.016529 \\ 
%14 & ASHL & 0.016425 \\ 
%15 & ADFR & 0.014804 \\ 
%16 & AWCR & 0.01388 \\ 
%17 & ADLR & 0.013676 \\ 
%18 & AWAL & 0.012399 \\ 
%19 & HSNL & 0.011773 \\ 
%20 & RMGL & 0.011302 \\ 
\hline 
\end{tabular} 
\begin{tabular}{|c|c|c|} 
\hline 
 & $\lambda_{10}=-0.45784$ & $|\psi^*_i|$ \\ 
\hline 
\hline 
1 & AVAL & 0.10651 \\ 
2 & AVAR & 0.079403 \\ 
3 & AVBR & 0.036779 \\ 
4 & VD05 & 0.025086 \\ 
5 & VA09 & 0.02438 \\ 
6 & VD06 & 0.020977 \\ 
7 & VA08 & 0.020242 \\ 
8 & AVBL & 0.019225 \\ 
9 & DD02 & 0.018684 \\ 
10 & PDB & 0.016485 \\ 
%11 & VA02 & 0.01642 \\ 
%12 & VD02 & 0.015503 \\ 
%13 & VB04 & 0.015487 \\ 
%14 & AS11 & 0.015451 \\ 
%15 & RID & 0.015056 \\ 
%16 & VD07 & 0.014528 \\ 
%17 & VA05 & 0.014055 \\ 
%18 & VB05 & 0.013819 \\ 
%19 & VB02 & 0.012662 \\ 
%20 & VD08 & 0.012634 \\ 
\hline 
\end{tabular}} 
\end{center}
\label{table2}
\end{table}

\begin{table}[!ht]
\caption{Same as in Table~\ref{table2}
for $48th$ largest eigenvalue modulus $|\lambda|$; 
 IPR are respectively $\xi \approx 54$ and $\xi \approx 47$.}
\begin{center}
\resizebox{7.5cm}{!}{
\begin{tabular}{|c|c|c|} 
\hline 
 & $\lambda_{48}=-0.30615-0.07037i$ & $|\psi_i|$ \\ 
\hline 
\hline 
1 & RIH & 0.017854 \\ 
2 & BDUR & 0.017737 \\ 
3 & OLLR & 0.016701 \\ 
4 & CEPDR & 0.016463 \\ 
5 & RMGR & 0.016357 \\ 
6 & AIAL & 0.016072 \\ 
7 & ASHR & 0.015585 \\ 
8 & VC04 & 0.015265 \\ 
9 & ASKR & 0.014 \\ 
10 & IL2R & 0.013978 \\ 
%11 & HSNR & 0.01237 \\ 
%12 & VC03 & 0.012326 \\ 
%13 & PVNL & 0.011528 \\ 
%14 & ALMR & 0.011124 \\ 
%15 & PVCL & 0.011079 \\ 
%16 & PVCR & 0.010816 \\ 
%17 & RIGR & 0.010807 \\ 
%18 & URBR & 0.010132 \\ 
%19 & AVBR & 0.0098765 \\ 
%20 & HSNL & 0.0096856 \\ 
\hline 
\end{tabular} 
\begin{tabular}{|c|c|c|} 
\hline 
 & $\lambda_{48}=0.26353-0.095716i$ & $|\psi^*_i|$ \\ 
\hline 
\hline 
1 & RMEV & 0.026461 \\ 
2 & RIBR & 0.013343 \\ 
3 & OLQDR & 0.013145 \\ 
4 & IL1DL & 0.012932 \\ 
5 & IL1DR & 0.012911 \\ 
6 & RIAR & 0.012896 \\ 
7 & RICR & 0.012728 \\ 
8 & OLQDL & 0.012586 \\ 
9 & RIGR & 0.012256 \\ 
10 & SMDDR & 0.011958 \\ 
%11 & RMDVL & 0.011617 \\ 
%12 & RMDDL & 0.010192 \\ 
%13 & RIGL & 0.0098548 \\ 
%14 & RMED & 0.0098161 \\ 
%15 & AIYL & 0.0095303 \\ 
%16 & AIBL & 0.0092849 \\ 
%17 & AVAR & 0.0089111 \\ 
%18 & RMDR & 0.0086201 \\ 
%19 & RMGR & 0.0083879 \\ 
%20 & OLQVL & 0.0083535 \\ 
\hline 
\end{tabular} 
}
\end{center}
\label{table3}
\end{table}

The top ten neurons in eigenfunction
amplitude for four specific eigenstates of $G$ and $G^*$ 
are given in Table~\ref{table2}, Table~\ref{table3}.
In Table~\ref{table2} we have eigenstates with low value of 
IPR so that the corresponding wavefunctions are 
localized essentially on only about 4 neurons
being $AIAR$, $AIAL$, $ASISL$, $ASIR$
and  $AVAL$, $AVAR$, $AVBR$
for $\lambda_{10}$ of $G$ and $G^*$ respectively.
In Table~\ref{table3} the values of IPR are rather large
and these eigenstates are
spread over many neurons.

\begin{figure}[!ht] 
\begin{center} 
\includegraphics[width=7.5cm]{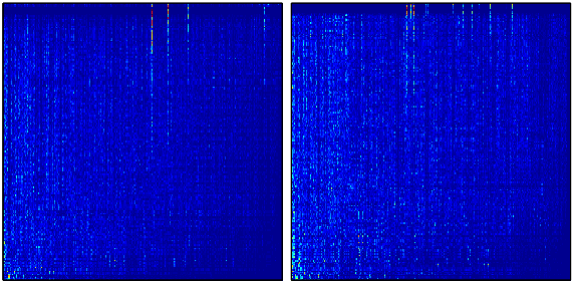}
\caption {(Colour on-line)
Dependence of amplitude of eigenstates  $|\psi_i(K)|$ of $G$
and  $|\psi^*_i(K^*)|$ of $G^*$ on PageRank index $K$
(left panel) and 
CheiRank index $K^*$ (right panel); here $x$-axis
shows values of $K$ and $K^*$, while $y$-axis
shows index $i$ of eigenstates
being ordered in a decreasing order of $|\lambda_i|$
(see text). The whole index range $1 \leq K, K^* \leq 279$ 
is shown with PageRank (CheiRank) vector being 
at the bottom line of each panel. 
The color is proportional to $|\psi_i(j)|$ changing 
from minimum blue value to maximum value in red.}
\label{fig6} 
\end{center}
\end{figure}

To determine if some eigenvectors are localized
on a certain group of neurons, we plot in Fig.~\ref{fig6} 
the amplitude of each eigenstate 
horizontally in the basis of neurons ordered by 
indexes of $K$ and $K^*$
of PageRank and CheiRank vectors.
The eigenstates of $G$ matrix  show 
four distinct vertical stripes at $K=149, K=165, K=185, K=261$ 
which correspond respectively 
to neurons $PVDR$, $IL2DR$, $IL2DL$, $PLNR$. 
The same plot for $G^*$ matrix shows a larger number of stripes
which have less pronounced amplitudes. 
These stripes of $G^*$ are located on the following neurons 
$K^*=116 ( RIPL)$, $K^*=123 (RIPR)$,  
$K^*=120 (AS07)$, $K^*=122 (AS10)$, 
$K^*=135 (DB06)$, $K^*=137 (DB05)$, 
$K^*=215 (DA07)$, 
$K^*=162 (VA10)$, $K^*=172 (SIADL)$, 
$K^*=181 (SIAVL)$, 
$K^*=199 (SIAVR)$, 
$K^*=221 (SIADR)$.

We think that an identification of eigenstates with
certain physiological functions of worm 
can be an interesting task which however requires
further more detailed studies in collaboration
with physiologists. The tables of top
20 nodes of eigenstates with 50 largest 
$|\lambda_i|$ values are available at
\cite{wormgooglematrix}.

\section{Discussion}

In this Letter, we  analyzed the neural network 
of  C.Elegans using Google matrix approach to directed networks
which proved its efficiency for the WWW.
We classify worm neurons using 
PageRank and CheiRank probabilities 
corresponding to 
the principal vectors
of the Google matrix with direct and inverted links. 
Thus neurons in the head region take top positions in
PageRank, CheiRank and combined 2DRank.
Also interneurons occupy 
top ranking positions. 
We show that influences and interdependency 
between neurons can be studied using the ImpactRank
propagator. We argue that the eigenvectors
with large modulus of eigenvalues of the Google matrix
may select specific physiological functions.
This conjecture still need to be investigated in more
detailed studies. 
Our research shows that
the Google matrix analysis 
represents a useful and powerful method of
neural network analysis.

%\section{ACKNOWLEDGMENT} 
We thank  Emma K.\ Towlson and Petra E.\ V\'ertes
for useful discussions and for providing us 
the links between neurons
available from C.elegans neural network 
data set at \cite{wormatlas}.
This work is supported in part by 
EC FET Open project "New tools and algorithms 
for directed network analysis" (NADINE No 288956).

\end{document}